\documentclass[12pt,preprint]{aastex}

\usepackage{amsmath}
\usepackage{array}
\usepackage{wasysym}
\usepackage{graphicx, subfigure}

\shorttitle{Late-forming Chondrule in Comet Wild 2}
\shortauthors{Ogliore et al.}

\begin{document}


\title{Incorporation of a Late-forming Chondrule into Comet Wild 2}


\author{R. C. Ogliore\altaffilmark{1}, G. R. Huss\altaffilmark{1}, and K. Nagashima\altaffilmark{1}}
\affil{$^{1}$Hawai\textquoteleft i Institute of Geophysics and Planetology, University of Hawai\textquoteleft i at M\={a}noa, Honolulu, HI 96822}

\and

\author{A. L. Butterworth\altaffilmark{2}, Z. Gainsforth\altaffilmark{2}, J. Stodolna\altaffilmark{2}, and A. J. Westphal\altaffilmark{2}}
\affil{$^{2}$Space Sciences Laboratory, University of California at Berkeley, Berkeley, CA 94720}

\and

\author{D. Joswiak\altaffilmark{3}}
\affil{$^{3}$Department of Astronomy, University of Washington, Seattle, WA 98195}

\and

\author{T. Tyliszczak\altaffilmark{4}}
\affil{$^{4}$Advanced Light Source, Lawrence Berkeley National Laboratory, Berkeley, CA 94720}

\begin{abstract}
We report the petrology, O isotopic composition, and Al-Mg isotope systematics of a chondrule fragment from the Jupiter-family comet Wild 2, returned to Earth by NASA's Stardust mission. This object shows characteristics of a type II chondrule that formed from an evolved oxygen isotopic reservoir. No evidence for extinct $^{26}$Al was found, with ($^{26}$Al/ $^{27}$Al)$_0$ $<$ 3.0$\times$10$^{-6}$. Assuming homogenous distribution of $^{26}$Al in the solar nebula, this particle crystallized at least 3 Myr after the earliest solar system objects---relatively late compared to most chondrules in meteorites. We interpret the presence of this object in a Kuiper Belt body as evidence of late, large-scale transport of small objects between the inner and outer solar nebula. Our observations constrain the formation of Jupiter (a barrier to outward transport if it formed further from the Sun than this cometary chondrule) to be more than 3 Myr after calcium-aluminum-rich inclusions.
\end{abstract}


\section{Introduction}
High-temperature objects---fragments of chondrules and calcium-aluminum-rich inclusions (CAIs)---have been discovered in the collection of samples returned by NASA's Stardust mission from comet Wild 2 (e.g. \citet{nak08}, \citet{mat10}). The diversity of the Wild 2 samples shows that this comet does not consist entirely of unaltered nebular material rich in presolar grains and interstellar amorphous silicates \citep{bro06}, nor is comet Wild 2 close in composition to a known class of meteoritic material or chondritic-porous interplanetary dust particles (e.g. \citet{ogl10}). The presence of chondrules and CAIs in a Jupiter-family comet requires a transport mechanism between the hot inner solar system, where these objects likely crystallized, and the scattered disk, where they were eventually accreted into the comet \citep{dun04}.

Large-scale transport in the disk must have occurred before the formation of Jupiter because the growing embryo of Jupiter strongly suppressed outward transport across its orbit \citep{cie09}. A nascent Jupiter efficiently accreted material that diffused into its orbital gap, effectively creating a barrier to outward transport of high-temperature objects if its orbit was farther from the Sun than where these objects formed.

Measurements of the decay products of short-lived radionuclides in high-temperature objects in the Stardust collection will yield an earliest time when this large-scale transport occurred. It is then possible to constrain the timing of the formation of Jupiter, which because of its large mass controlled much of the dynamics of planetesimals forming in the young solar system (e.g \citet{wal11}).

\section{Sample Preparation}

Using synchrotron X-ray fluorescence and X-ray absorption near-edge spectroscopy, we measured the elemental composition and Fe oxidation state of $\sim$200 fragments in 11 Stardust tracks in millimeter-sized aerogel keystones (see \citet{ogl10} and references therein). One of the terminal particles in Track C2052,2,74 appeared to contain a complex assemblage of olivine grains as well as Cr- and Ni-rich phases. This cometary particle, named ``Iris'', was chosen for extraction and further analysis; it is 23$\times$10$\times$15$\mu$m in size. 

The keystone containing Track C2052,2,74 was trisected into three blocks using glass needles robotically controlled by micromanipulators.  The block containing Iris was again segmented with glass needles into a thin wafer $\sim$100 $\mu$m thick, embedded in epoxy, and attached to the end of an epoxy bullet.  Approximately 25\% of Iris was then ultramicrotomed into slices $\sim$100 nm thick and placed on transmission electron microscopy (TEM) grids with 10 nm amorphous carbon substrates.  These microtome slices were analyzed by synchrotron-based scanning transmission X-ray microscopy (STXM) and TEM. Approximately 75\% of Iris remained in the epoxy bullet, its cross section exposed (a ``potted butt'') for O isotope measurements.

\section{Mineralogy/Petrology}
\begin{figure}[h]
       \centering
        \includegraphics[width=\textwidth]{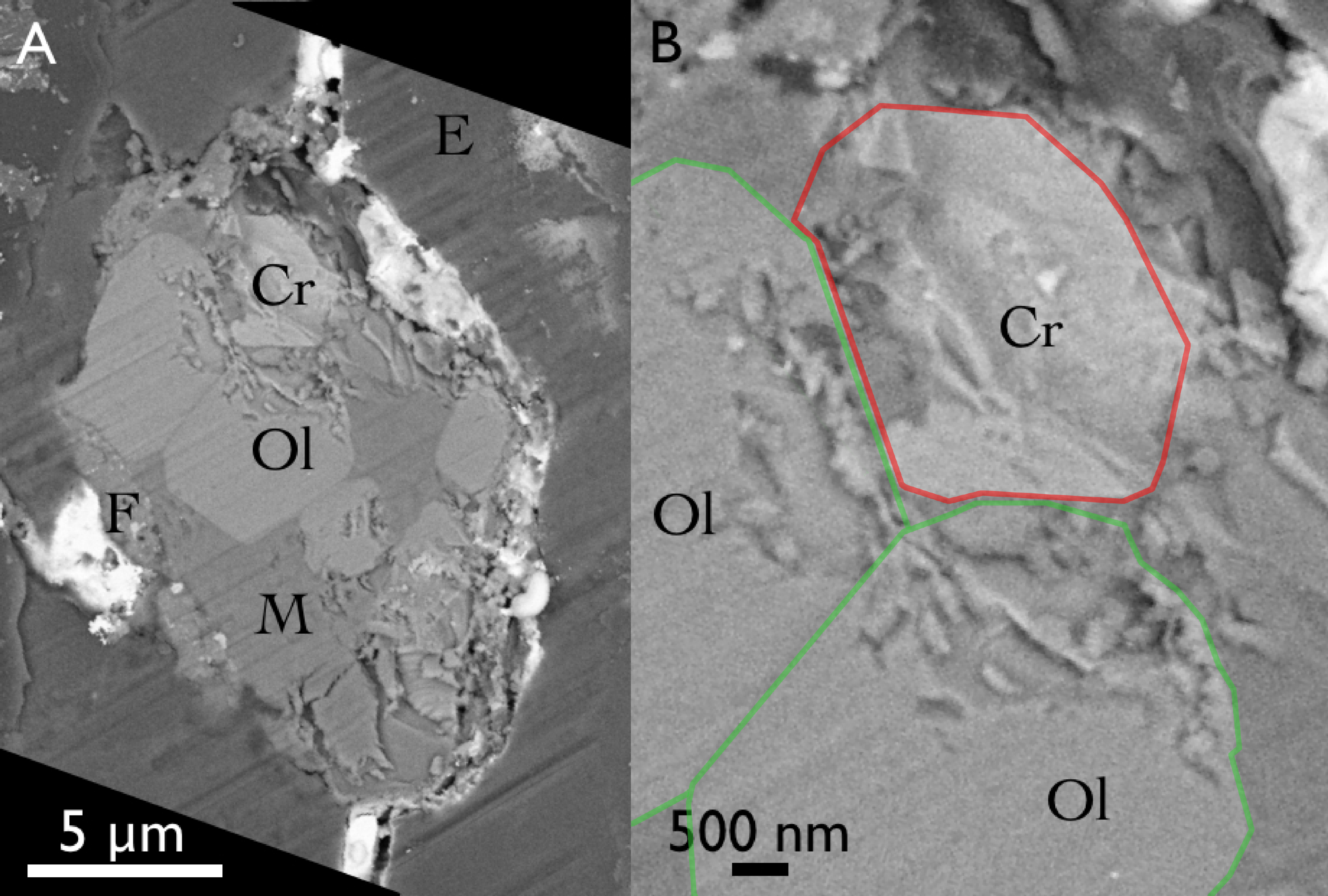}
        \caption{Backscatter electron image of the Iris potted butt, showing (a) the exposed face of the grain and phases (Ol: olivine; Cr: chromite; M: mesostasis; F: fusion crust; E: epoxy+aerogel) and (b) a closer view of the compact interface between two olivine grains and chromite with the olivines outlined in green and the chromite outlined in red. \label{fminpet}}
 \end{figure}

Iris contains a complex assemblage of grains with at least nine distinct crystals embedded in an SiO$_2$-rich glassy mesostasis containing Al, Na, Ca, Fe, and Mg in decreasing order of abundance.  The mesostasis has some crystalline order and could contain plagioclase but it is not fully crystalline; the mesostasis is also heterogeneous showing a $\sim$50\% variation in Al, Mg, and Fe content over a distance of $<$10 $\mu$m. The crystals are Fe-rich olivines (nearly equilibrated, ranging from Fo$_{60}$ to Fo$_{68}$) with varying minor amounts of Ca, Mn, and Al. Additionally, there are multiple spinels in Iris which are enriched in Cr and Fe (one spinel is Chr$_{61}$Sp$_{34}$) with minor Ti, V and Mn, and sodic plagioclase (An$_{14\text{--}22}$).  Two plagioclase crystals have been verified as oligoclase via diffraction.  Olivine crystals have compact interfaces with chromite and appear to have co-crystallized (Fig. \ref{fminpet}(b)). Elemental
quantifications for an olivine in the potted butt and a chromite and mesostasis in a TEM grid are given in Table \ref{tab:ElementalQuants}. 

Using the olivine-spinel thermometer we find an equilibration temperature $>$1100$^{\circ}$C for Iris, signifying rapid cooling \citep{sac91}.  The exchange partitioning of Fe and Mg between olivine and spinel was measured as K$_{\textnormal{\scriptsize D}}$=(Fe/Mg)$^{\textnormal{\scriptsize sp}}$/(Fe/Mg)$^{\textnormal{\scriptsize oliv}}$ $ \approx$ 4.7.  Comparison of K$_{\textnormal{\scriptsize D}}$ and Cr, Al content in the chromite with type II chondrules from ordinary and carbonaceous chondrites \citep{joh91} shows that this particle experienced little metamorphic alteration---most likely less than 3.2 on the petrologic scale for chondrites.  The compositions of the minerals along with their textural relationships and the evidence for high temperature diffusion and rapid cooling strongly resembles type II chondrules seen in chondritic meteorites.  Chondrules are expected to be spherical in shape, testifying to their once molten state. However, aerogel capture can alter the shapes of cometary grains captured by Stardust. The overall shape of Iris is ellipsoidal and cross-sectional views show it to have a porphyritic structure (Fig. \ref{fminpet}(a)).  However, its pre-capture shape and size are not known, so we cannot determine if it is a primary object or a fragment of a type II chondrule.   We conclude that Iris has experienced a relatively simple igneous history.

\section{O isotope measurements}

To acquire precise O-isotope data for Iris, we developed a sample mount consisting of an Al round, which supported the potted butt at a fixed altitude, covered by an Al annulus which both supported a 500nm-thick membrane (made of gold-coated Si$_3$N$_4$ with a 300$\mu$m ion-milled hole) and incorporated polished standards surrounding the central hole. San Carlos olivine was used to standardize Iris olivine, and Miyakajima plagioclase and Burma spinel were used to standardize Iris mesostasis and chromite, respectively. 

We measured three oxygen isotopes in Iris using the University of Hawai\textquoteleft i Cameca ims 1280 ion microprobe in multicollection mode ($^{16}$O on a Faraday cup, $^{17}$O and $^{18}$O on electron multipliers). A 25--30 pA Cs$^{+}$ primary ion beam was focused to $\sim$2 $\mu$m to allow for single-grain analysis of Iris. The data was corrected for background, deadtime, detector yield, and interference from $^{16}$OH$^{-}$ to $^{17}$O (typically $<$0.2$\permil$). We corrected measured compositions of the mineral phases in Iris for instrumental mass fractionation by comparing with appropriate mineral standards of known composition.

We measured the oxygen isotopic composition of San Carlos olivine mounted analogously to Iris in order to understand instrumental mass fractionation associated with the mounting of the unknown sample relative to the standards. We found that there was a reproducible instrumental fractionation effect between the San Carlos olivine surrounding the central hole and the San Carlos olivine in the central hole (the center olivine plotted $\sim$3$\permil$ lower in $\delta^{18}$O than the surrounding olivine along the terrestrial fractionation line). We applied this offset to the final Iris oxygen measurements. Our conservative estimate of the uncertainties in the final ratios comes from adding in quadrature the statistical uncertainty of the individual measurements, the standard deviation of the standard measurements, and the uncertainty in the instrumental mass fractionation offset.

Olivine, chromite, and mesostasis in Iris have oxygen isotopic compositions indistinguishable from terrestrial oxygen (Figure \ref{oiso}).  The oxygen isotopic compositions of the three mineral phases are consistent with co-crystallization from a single melt. Iris has oxygen-isotope composition higher in both $^{18}$O/$^{16}$O and $^{17}$O/$^{16}$O than most chondrules in meteorites and other chondrule-like objects in the Stardust collection (Figure \ref{oiso}). The oxygen isotopes of nebular solids can evolve toward isotopically heavier compositions through processes such as interaction with water (partitioning between liquid and solid) or evaporation (Rayleigh-like distillation). Iris apparently formed from a relatively evolved oxygen isotopic reservoir.

\section{Al--Mg isotope measurements}

\begin{figure}[h]
       \centering
        \includegraphics[width=\textwidth]{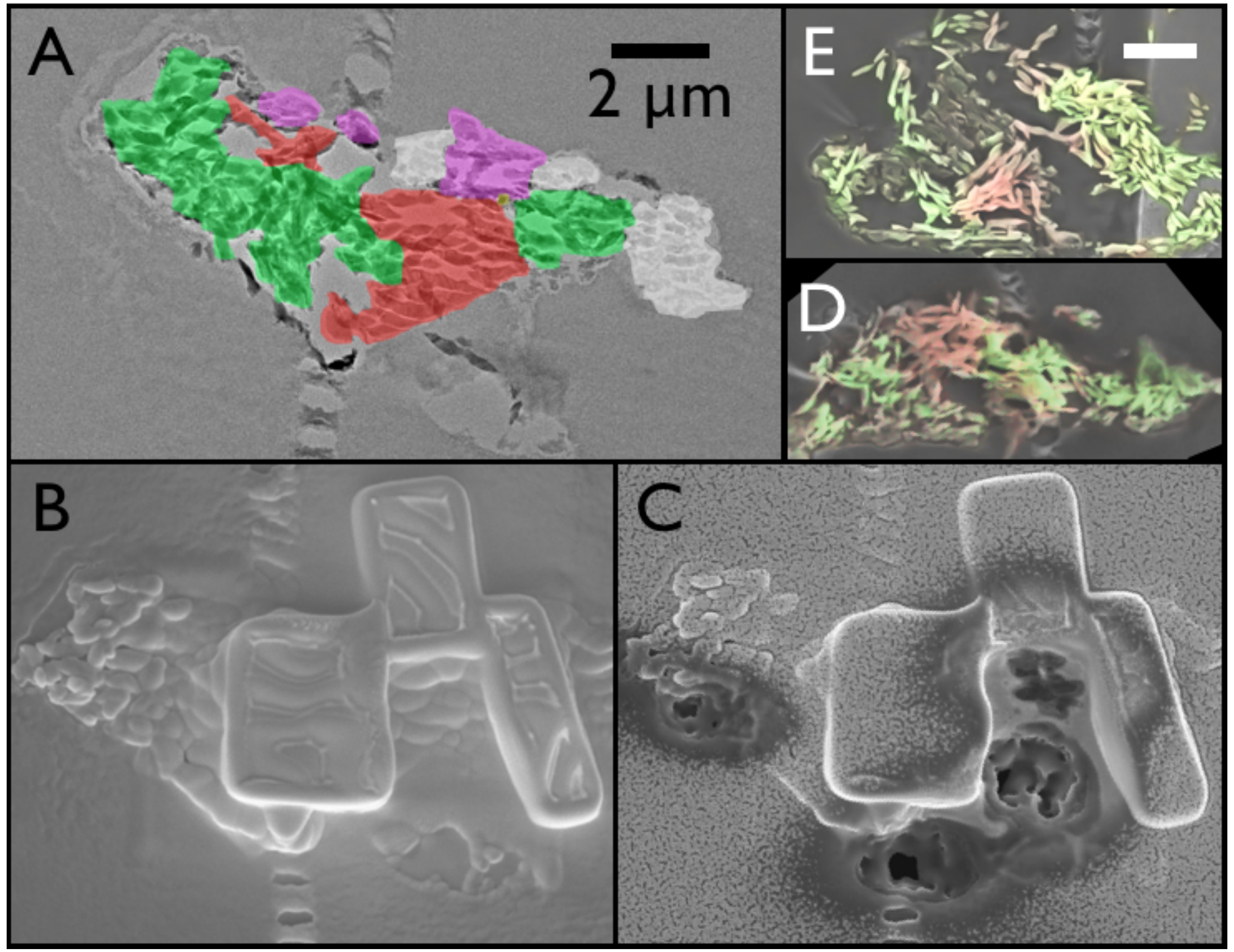}
        \caption{TEM and scanning electron microscopy (SEM) images of the Iris TEM grid used for Al-Mg isotope measurements.  (a) TEM bright-field image with colored overlays showing the phase containing the highest Al/Mg ratio (red) with oligoclase composition as well as olivine (green), Si-rich glass with Al, Mg and Fe (purple), SiO$_{2}$ glass (white) and a small iron sulfide (yellow). (b) The same section after masking with Pt to shield the Mg-rich olivine and glass.  (c) Same, after isotopic measurements were complete.  (d) and (e) LRGB energy-dispersive X-ray spectroscopy (EDS) maps showing two additional sections measured on the same grid before masking.  Al is red, Mg is green, and a secondary electron image provides the luminance channel. Both scale bars are 2 $\mu$m. \label{gridfig}}
 \end{figure}

We identified a crystalline phase of oligoclase composition in four ultramicrotomed slices of Iris on a single grid by TEM. This is an Al-bearing phase with only trace Mg, making it appropriate to search for evidence of extinct $^{26}$Al (t$_{1/2}$ $\approx$ 0.73 Myr), which can be used to date Iris relative to other meteoritic material. The TEM grid was prepared for Al--Mg isotopic measurements in the Cameca ims 1280 by masking nearby Mg-rich phases with Pt deposited by focused ion beam (FIB) and gluing the grid to an Al stub with graphite paint (Figure \ref{gridfig}). Terrestrial anorthite and olivine standards were prepared using the same methodology without FIB masking.

Using a primary $^{16}$O$^{-}$ beam at 3--10 pA with $\sim$2 $\mu$m beam-spot size and $\sim$4800 mass-resolving power, we made simultaneous measurements of Mg isotopes followed by peak-jumping of the magnetic field to $^{27}$Al. We measured the Iris oligoclase until the sample was almost entirely sputtered away (50--100 cycles with 15 s per cycle for Mg isotopes, 1 s for $^{27}$Al). We estimated the effects of instrumental mass fractionation from measurements of the terrestrial anorthite standard and found no significant fractionation. 

We assumed that the expected value of $^{26}$Mg counts in each cycle of the Iris measurement is equal to the sum of nonradiogenic $^{26}$Mg and radiogenic $^{26}$Mg. The nonradiogenic component is equal to the counts of $^{24}$Mg multiplied by a linear instrumental mass fractionation correction (determined by the standards as described previously) and a sample linear mass fractionation correction (a free parameter). The radiogenic component is equal to the $^{27}$Al counts (corrected for the different count times between the Mg isotopes and $^{27}$Al) multiplied by ($^{26}$Al/$^{27}$Al)$_0$ (a free parameter). Additionally, the nonradiogenic component of the expected $^{26}$Mg counts in a cycle should be proportional to the measured $^{25}$Mg counts. We assumed that the counts in each cycle follow a Poisson distribution with mean given by the expectation value described above with either $^{24}$Mg or $^{25}$Mg as the natural reference isotope, and maximized the log-likelihood function of all the measured cycles with both reference isotopes simultaneously. The maximum of the log-likelihood function occurs at the most likely values of the sample linear mass bias and ($^{26}$Al/$^{27}$Al)$_0$ as described by our collected data. We determined robust confidence intervals on these two parameters by Monte Carlo simulation with 10$^5$ trials, assuming statistical and systematic uncertainties of isotope counts, isotope ratios, and instrumental mass fractionation as determined from our standard measurements.

We found no evidence of extinct $^{26}$Al in Iris. Our one-sided 2$\sigma$ upper bound on ($^{26}$Al/$^{27}$Al)$_0$ in Iris is 3.0$\times$10$^{-6}$. Assuming an $^{26}$Al/$^{27}$Al ratio of  5.0$\times$10$^{-5}$ when CAIs formed \citep{mac95}, which were the first solar system solids, Iris formed at least 3.0 Myr after CAIs. However, the level of heterogeneity of $^{26}$Al in the solar nebula is uncertain (e.g. \citet{vil09} report homogeneity of $^{26}$Al and Mg isotopes at $\pm$10\% while \citet{lar11} claim up to 80\% heterogeneity in the initial abundance of $^{26}$Al in the inner solar system). If we assume instead that the material from which Iris formed had the lowest ($^{26}$Al/$^{27}$Al)$_0$ in chondrule-containing parent bodies (from \citet{lar11}), the ordinary chondrite parent body, ($^{26}$Al/$^{27}$Al)$_0$=1.63$\times$10$^{-5}$, then Iris crystallized at least 1.8 Myr after CAIs. 

\begin{figure}[h]
       \centering
        \includegraphics[width=\textwidth]{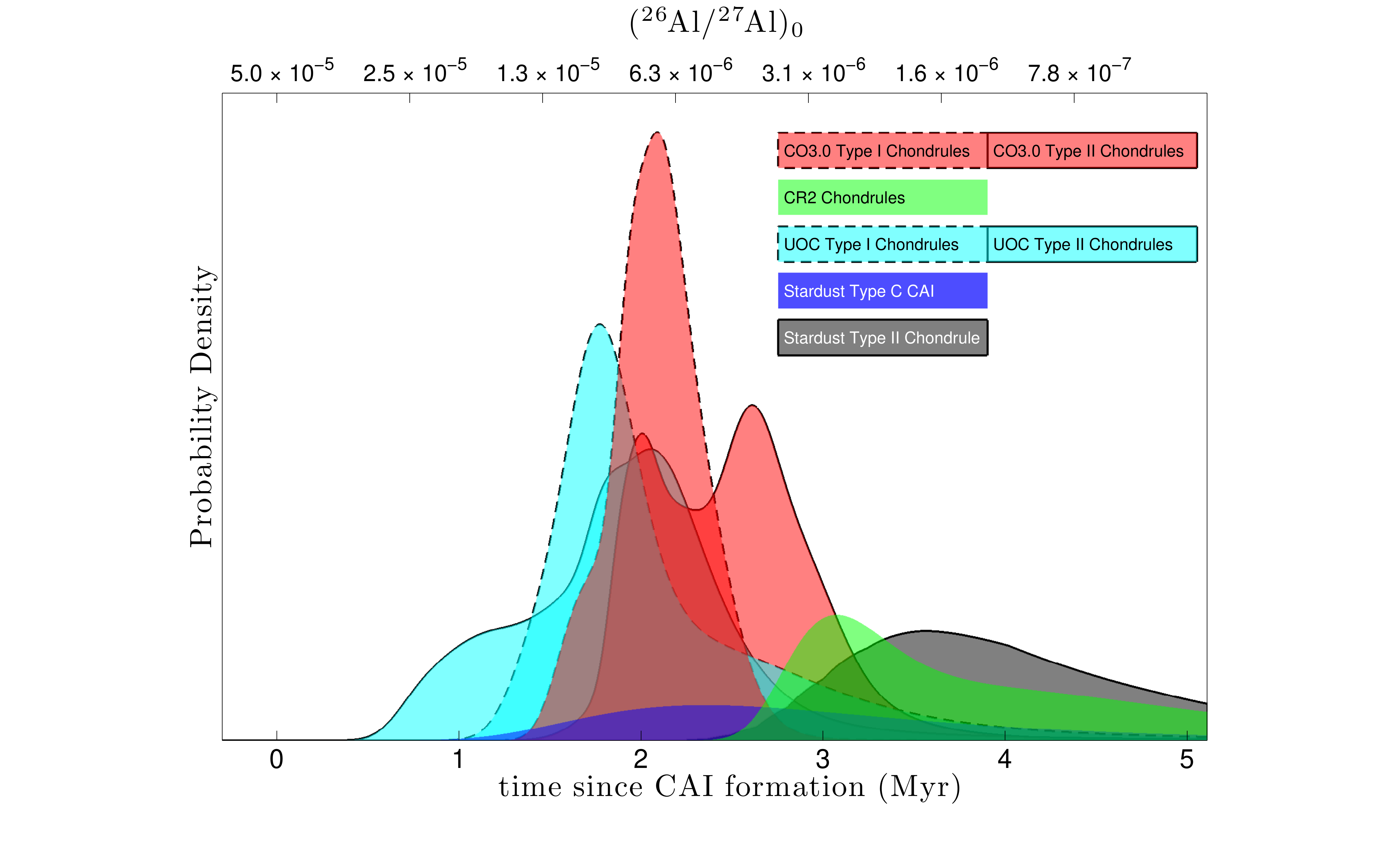}
        \caption{Probability density distributions for the formation times of type I and type II chondrules from CO3.0, CR2, and UOC meteorites compared to the type II Stardust chondrule fragment ``Iris'' (this work) and the Stardust Type C CAI fragment ``Coki'', all from SIMS measurements of the Al-Mg isotope system. The probability density curve (arbitrary units) is derived from the sum of measurements with quoted normal uncertainties for the meteorite classes: 12 type I and 14 type II CO3.0 chondrules \citep{kur08,kun04}, 6 type I and 2 type II CR2 chondrules \citep{nag08} combined into one distribution , and 3 type I and 26 type II UOC chondrules \citep{rud07,kit00,mos02}. The Stardust Type C CAI curve is from a single object \citep{mat10}. The probability density curve shown for Iris is derived from the maximum likelihood and Monte Carlo technique employed to calculate initial ($^{26}$Al/$^{27}$Al). Homogeneity of solar nebula $^{26}$Al and a canonical initial ($^{26}$Al/$^{27}$Al) of 5$\times$10$^{-5}$ was assumed for CAI formation. Probability density longer than seven half-lives of $^{26}$Al and probability density contribution from negative isochron slopes are not shown. \label{almg}}
         \end{figure}

The formation time of Iris compared to chondrules from other meteorites, as well as a CAI-like Stardust fragment, is shown in Figure \ref{almg}. The probability density curve for Iris is derived from the Monte Carlo simulations discussed above, whereas the curves for other objects are normalized sums of reported measurements of ($^{26}$Al/$^{27}$Al)$_0$, assuming Gaussian-distributed errors. Probability density associated with negative ($^{26}$Al/$^{27}$Al)$_0$ corresponds to ill-defined time since CAI formation, so these densities, as well as those for $>$7 half-lives of $^{26}$Al, are not shown. (Consequently, the area under the curves for objects with allowed negative or very small ($^{26}$Al/$^{27}$Al)$_0$ is less than for objects that formed earlier.)

\section{Discussion}
Since CAIs are believed to have formed when the Sun was a class 0 or class I protostar, Iris formed late in the evolution of the solar nebula, at a time when $\sim$95\% of the scattered disk is thought to have cleared \citep{her01}. The residual disk material may have coated Jupiter-family comets with a ``late veneer'' enriched in inner solar system material \citep{ogl10}, and this may explain significant differences between the Stardust sample, which sampled Wild 2 coma material ejected by jets that entrained near-surface dust \citep{bel10}, and chondritic porous  interplanetary dust particles, which probably sample the bulk of Kuiper Belt comets \citep{nes10}.  

\begin{figure}[h]
       \centering
        \includegraphics[width=\textwidth]{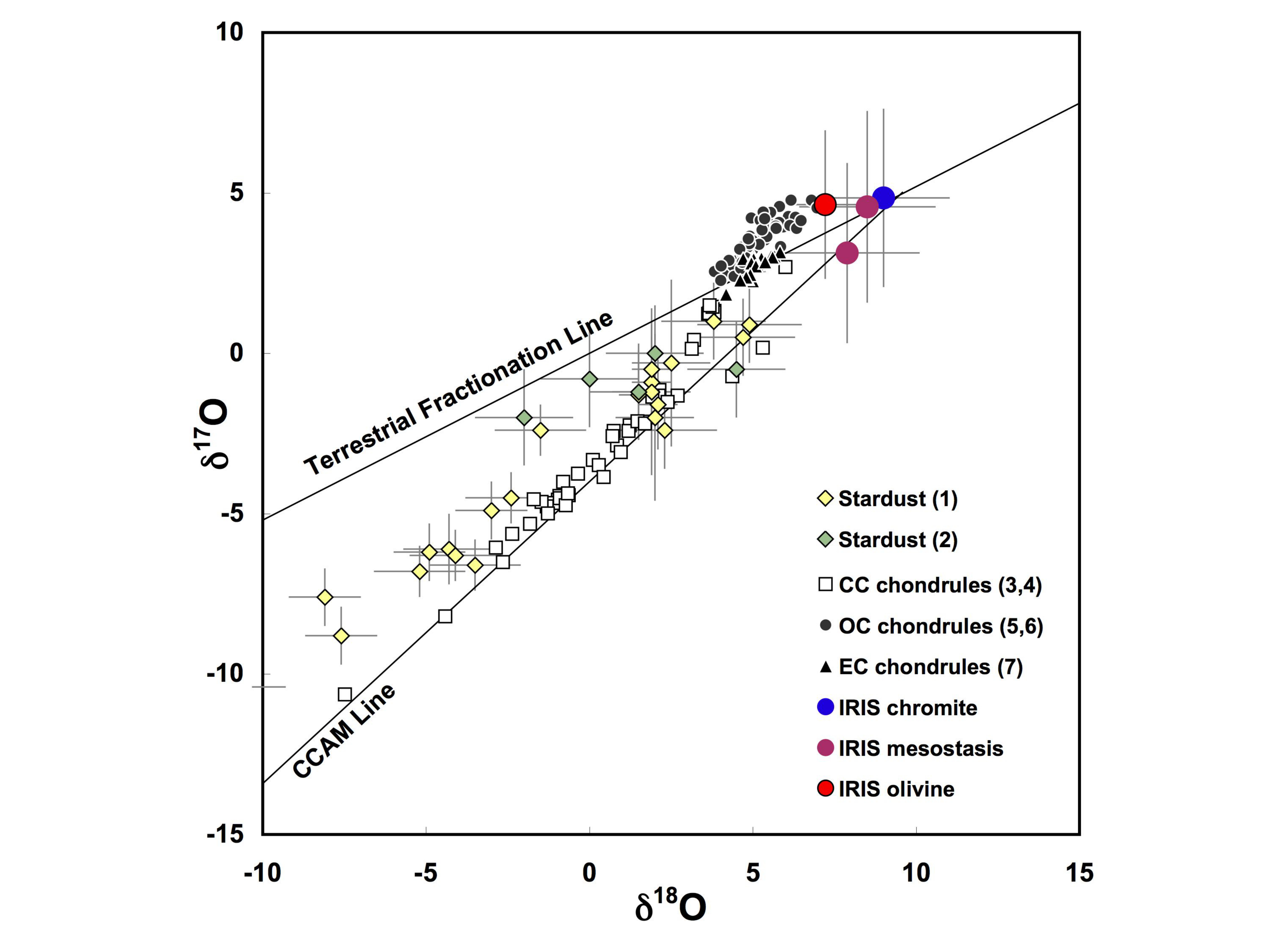}
       \caption{Oxygen isotopic composition of phases in the Stardust type II chondrule fragment Iris compared with previous O isotope measurements of chondrule-like Stardust particles (1: \citet{nak08}, 2: \citet{mck06}) and bulk chondrules from carbonaceous (CC, (3: \citet{cla83}, 4: \citet{jon04})), unequilibrated ordinary (OC, (5: \citet{cla91}, 6: \citet{bri98})), and enstatite (EC, (7: \citet{cla85}) chondrites. The quantities $\delta^{17}$O and $\delta^{18}$O are permil deviations from $^{17}$O/$^{16}$O and  $^{18}$O/$^{16}$O of Standard Mean Ocean Water, respectively. The terrestrial fraction line and carbonaceous chondrite anhydrous mineral (CCAM) line are denoted in the figure.  \label{oiso}}
 \end{figure}

These analyses are for a single object in the Stardust collection. However, Iris is not unique or even particularly unusual in the suite of particles returned from comet Wild 2. \citet{nak08} identified four chondrule-like objects in the Stardust samples which are more $^{16}$O-enriched than Iris (Figure \ref{oiso}) and olivines that are less Fe-rich (Fo$_{79\text{--}80}$, Fo$_{95}$, Fo$_{91}$).  At least one chondrule fragment was also identified during the Stardust preliminary examination \citep{zol06}. All of these objects show igneous textures similar to Iris, though Iris appears to have formed in a more oxidizing environment. Recent measurements by \citet{jos11} show two fragments of Fe-rich olivine (Fo$_{62\text{--}67}$, Fo$_{58\text{--}61}$), that also have O isotopic composition close to Iris. Chondrule-like objects from Stardust show a broad range of isotopic and mineralogical compositions; a subset of these objects could be genetically related to Iris.

The formation time of Iris is most consistent with the late-forming chondrules in CR chondrites. Chondrules from CB chondrites also formed relatively late (as determined by $^{207}$Pb--$^{206}$Pb measurements), likely by a giant impact in the early solar system \citep{kro05}. The parent bodies of both CR and CB chondrites were scarcely heated \citep{sco06}, similar to Wild 2 \citep{bro06}. However, type II chondrules like Iris (and similar Stardust fragments) are very rare in CR and CB meteorites: $<$1\% of all chondrules \citep{wei93,kro02}. The Iris mesostasis is Fe-poor and Na-rich compared to mesostasis in type II chondrules in unequilibrated CR chondrules, and the Iris olivine is Ca- and Al-rich relative to olivine in CR type II chondrules \citep{ber11}. Although Iris is similar to chondrules from CR and CB meteorites in that it formed relatively late, Iris and similar Stardust fragments are unlikely to have originated from the CR- or CB-chondrite-forming region.

The Fe-rich olivines in Iris must have formed in an environment with oxygen fugacity higher than the typical redox conditions of the steady-state solar nebula \citep{kro00}. Shocks in the outer solar nebula, beyond the water-snow line at $\sim$5 AU \citep{cyr98}, were rich in water vapor \citep{cie03} which could have formed a chondrule like Iris near the current orbit of Jupiter. However, the predicted composition of chondrule olivine in such a shock is strongly peaked between Fo$_{76}$ and Fo$_{89}$, Fo$_{70}$ olivine and lower is predicted to be $\sim$3\% \citep{fed08} of all chondrule olivine generated in the shock. Therefore we conclude that Iris was probably not created in an outer-nebula shock, but likely formed in the inner solar nebula, from material with relatively high Fe/Mg or in a region of high oxygen fugacity \citep{jon90}, more than 3 Myr after CAIs.

Recent measurements of oxygen isotope variations in the rim of a CAI from the CV3 chondrite Allende \citep{sim11} indicate that these early objects experienced circulation in the solar nebula. A Type C CAI found in the Stardust samples was constrained to have crystallized at least 1.7 Myr after the onset of CAI formation (assuming a homogeneous nebular reservoir of canonical ($^{26}$Al/$^{27}$Al)$_0$), though it likely experienced a complex history \citep{mat10}. Additionally, high-temperature components (CAIs and chondrules) do not appear to be scarce in the Stardust samples. These observations provide strong evidence of a dynamic early solar system transporting material between the inner and outer nebula.

The formation of Iris in the inner nebula requires it to be transported to the scattered disk at $\sim$35 AU \citep{dun04,tir09} where it was incorporated into comet Wild 2.  This transport could have occurred outside the plane of the disk (e.g. \citet{cie07}), by diffusion (e.g. \citet{cuz03}), aerodynamic lofting and radial drift (e.g.  \citet{cie08}), or outward advective flows (e.g. \citet{hug11}). Jupiter's growing embryo would open a gap in the disk \citep{bat03} which would prohibit outward transport \citep{cie09} if Iris was created in an inner-nebula shock (unless Jupiter had migrated inside of 2 AU and was closer to the Sun than Iris when Iris formed \citep{wal11}). Iris likely formed in an event prior to Jupiter's formation, such as in an inner-nebula spiral shock that existed before Jupiter opened a gap in the disk \citep{bol05}. Therefore, our measurements set a constraint on the formation time of Jupiter of at least 3 Myr after CAI formation making it unlikely that Jupiter formed early (e.g., by disk instability \citep{bos01}). 

Our constraint on the formation time of Jupiter is consistent with arguments based on the requirement to accrete asteroids (with constituent radiometric-dated chondrules and CAIs) before Jupiter grows large enough to inhibit accretion ($>$3--5 Myr, \citet{sco06}). Jupiter was estimated to form $\sim$3.3 Myr after the onset of planetesimal fragmentation in the main belt by \citet{bot05}, also consistent with the outward transport of Iris. The formation of chondrules in planetesimal bow shocks caused by Jovian resonances \citep{wei98} requires Jupiter to form $\sim$1 Myr after CAIs, a scenario which is disallowed by our measurements.

\acknowledgments
This work was supported by NASA grant NNX07AM62G (G.R.H.) and NNX07AM67G (A.J.W.). The operations of the Advanced Light Source and National Center for Electron Microscopy at Lawrence Berkeley National Laboratory are supported by the Director, Office of Science, Office of Basic Energy Sciences, U.S. Department of Energy under contract number DE-AC02-05CH11231. The authors thank the anonymous reviewer for helpful suggestions. Iris was named by Laura Westphal.

\clearpage
\newpage

\begin{table}
\begin{tabular}{|>{\centering}p{1in}|c|c|c|c|c|c|c|c|c|c|}
\hline
Sample & SiO$_{\text{2}}$ & Na$_{\text{2}}$O & MgO &
Al$_{\text{2}}$O$_{\text{3}}$ & CaO & TiO$_{\text{2}}$ &
V$_{\text{2}}$O$_{\text{3}}$ & Cr$_{\text{2}}$O$_{\text{3}}$ & MnO &
FeO\tabularnewline
\hline
Fo$_{\text{64}}$ Olivine (Potted butt) & 36 & - & 31 & 0.14 & 0.5 & -
& - & - & 0.8 & 31\tabularnewline
\hline
Mesostasis (Grid B9) & 58--65 & 15--16 & 0--3 & 11--22 & 3--5 & - & - &
- & - & 0--3\tabularnewline
\hline
Chromite
(Grid B9) & - & - & 6 & 17 & - & 1.5 & 0.6 & 46 & 0.8 & 28\tabularnewline
\hline
\end{tabular}
\caption{Elemental quantifications for various phases given in oxide weight \%,
the location of the measurement is given in parentheses. In the case of
chromite, V$_{\text{2}}$O$_{\text{3}}$ and MnO were measured with
STXM, all other measurements are TEM or SEM EDS. For the mesostasis,
a range is given to delineate the variation in that phase.\label{tab:ElementalQuants}}
\end{table}

\clearpage

\clearpage

\clearpage

\clearpage




\clearpage

\clearpage



\clearpage




\end{document}